\documentclass[12pt,english,longbibliography,aps,superscriptaddress,footinbib,12pt,sort&compress]{article}
\usepackage{ae,aecompl}
\usepackage[T1]{fontenc}
\usepackage[latin9]{inputenc}
\usepackage[active]{srcltx}
\usepackage{color}
\usepackage{url}
\usepackage{amsmath}
\usepackage{graphicx}
\usepackage[numbers]{natbib}

\makeatletter
\usepackage{aecompl}\usepackage{epstopdf}

\usepackage{babel}
\addto\captionsenglish{}
\usepackage[labelfont=bf]{caption}
\usepackage{caption}
\makeatother

\usepackage{babel}
\begin{document}
\date{}
\title{ {Unraveling the dislocation core structure at a van
der Waals gap in bismuth telluride}}
\author{\noindent D. L. Medlin$^{1}$, N. Yang$^{1}$, C. D. Spataru$^{1}$,
L. M. Hale$^{2}$ and Y. Mishin$^{3*}$}
\maketitle
\noindent \begin{center}
$^{1}$ Sandia National Laboratories, Livermore, CA 94551, USA\\
$^{2}$ Materials Measurement Laboratory, National Institute of Science
and Technology, Gaithersburg, MD 20899, USA\\
$^{3}$ Department of Physics and Astronomy, MSN 3F3, George Mason
University, Fairfax, Virginia 22030, USA
\par\end{center}
\begin{abstract}
 {Tetradymite-structured chalcogenides such as bismuth
telluride (Bi$_{2}$Te$_{3}$) are of significant interest for thermoelectric
energy conversion and as topological insulators. Dislocations play
a critical role during synthesis and processing of such materials
and can strongly affect their functional properties. The dislocations
between quintuple layers present special interest since their core
structure is controlled by the van der Waals interactions between
the layers. In this work, using atomic-resolution electron microscopy,
we resolve the basal dislocation core structure in Bi$_{2}$Te$_{3}$,
quantifying the disregistry of the atomic planes across the core.
We show that, despite the existence of a stable stacking fault in
the basal plane gamma surface, the dislocation core spreading is mainly
due to the weak bonding between the layers, which leads to a small
energy penalty for layer sliding parallel to the van der Waals gap.
Calculations within a semidiscrete variational Peierls-Nabarro model
informed by first-principles calculations support our experimental
findings.}
\end{abstract}
\vfill{}

\noindent $^{*}$Correspondence and requests for materials should
be addressed to Y.~M.~(email: ymishin@gmu.edu).

\pagebreak 
\section*{Introduction}
Layered, tetradymite-structured chalcogenides are of tremendous
technological interest due to the novel electronic and thermal transport
properties that are imparted by their quasi-two-dimensional (2D),
sheet-like structures. In such structures, thin sheets composed of
several atomic layers weakly interact with each other by van der Waals
forces across interlayer regions called van der Waals gaps. These
materials have long been of great interest as thermoelectrics \citep{RN595,RN593,RN594,RN158,RN54,RN72,He:2018aa}
and, more recently, have been of intense focus in the context of topological
insulators \citep{RN585,RN440,RN597,RN596,Walsh:2017aa}. It is important
to understand the nature of extended crystallographic defects, such
as dislocations, in these materials. Dislocations are 1D topological
defects in crystalline materials whose glide in certain crystallographic
planes constitutes the main mechanism of plastic deformation \citep{Hirth}.
A typical dislocation consists of a core region with large atomic
displacements from perfect-lattice positions, and an elastic strain
field extending deep into the surrounding crystal lattice. Dislocations
in the layered chalcogenides are relevant to both the processing and
functional properties of these materials. Polycrystalline bulk thermoelectrics
have long been processed by thermomechanical means that introduce
dislocations through plastic deformation in order to improve their
densification and control their crystallographic texture \citep{Medlin2009}.
Dislocations are also relevant in the context of chalcogenide nanostructures
and epitaxial films, for which growth spirals and low-angle grain
boundaries associated with threading dislocations are commonly reported
\citep{RN588,Hao:2013aa,RN409}.

Dislocations in quasi-2D chalcogenides can also strongly affect functional
properties of these materials by several mechanisms. For example,recent
work has suggested that dislocations present at low angle grain boundaries
in Bi$_{2}$Te$_{3}$-based alloys effectively scatter phonons in
the mid-frequency range \citep{RN590,RN592}, providing a grain-boundary
design strategy for engineering new materials with improved thermoelectric
energy conversion efficiency. The core structure of the dislocations
affects their mobility under applied mechanical stresses and thus
the dislocation density in the microstructure, as well as the effectiveness
of phonon scattering by individual dislocations. The large strain
fields near dislocation cores can also affect the electronic band
structure. For instance, scanning tunneling microscopy measurements
near low angle grain boundaries in Bi$_{2}$Se$_{3}$ thin films grown
by the molecular beam epitaxy method have demonstrated shifts in the
energy of the Dirac state, which were attributed to the large strain
fields near the individually resolved dislocations cores \citep{RN588}.
Such strain fields are characterized by alternating tension and compression
regions. Experiments and first-principles calculations \citep{RN588,Yang:2018aa}
have shown that application of tensile and compressive strains across
the van der Waals gap in Bi$_{2}$Se$_{3}$ strongly affects the electronic
band structure. Tensile strain was found to shift the Dirac point
while compressive strain opened a gap and destroyed the Dirac states.
These observations suggest that electronic and other functional properties
of the material can be strongly impacted by the core structure of
dislocations present at the van der Waals gap. The motion and interaction
of dislocations under thermomechanical processing can also strongly
alter the carrier concentrations, which directly affects properties
such as the electrical resistivity and thermopower \citep{Hu:2018aa,Jung:2018aa,Schultz:1962aa}.

A central question that remains unresolved is how the weak bonding
across the van der Waals gap affects the atomic structure of the dislocations
present at the interlayer. This is important since the structure of
a dislocation is central to its properties and behavior. For instance,
the local elastic strain distribution as well as the specific atomic
arrangements in the vicinity of a dislocation core depend on whether
the core is compact or dissociated into partial dislocations, which
directly impacts the local electronic states and behavior of charge
carriers \citep{Reiche:2016aa,Kweon:2016aa}. In this paper, we determine
 the structure of dislocations present at the basal plane of Bi$_{2}$Te$_{3}$,
a prototypical tetradymite-structured compound. Our approach is a
combination of experimental observations employing atomic-resolution
electron microscopy, and computer modeling encompassing \emph{ab initio}
electronic structure calculations and continuum/discrete dislocation
theory. The local structural disruption imposed by a dislocation presents
us with an opportunity to probe the interaction strength across the
interlayer gap. Our measurements of the dislocation core spreading,
analyzed using a semidiscrete Peierls-Nabarro framework, provide fundamental
insights concerning the dislocations in Bi$_{2}$Te$_{3}$ and, by
extension, other layered materials. This work additionally allows
us to compare the predictive capabilities of several commonly used
exchange-correlation functionals incorporating van der Waals corrections
into the density functional theory (DFT) calculations. The approaches
described here are generally applicable and provide a path to a deeper
understanding of extended defect structures and their connection to
interlayer bonding in this important class of materials.

\section*{Results}

\begin{figure}

\includegraphics[width=0.95\textwidth]{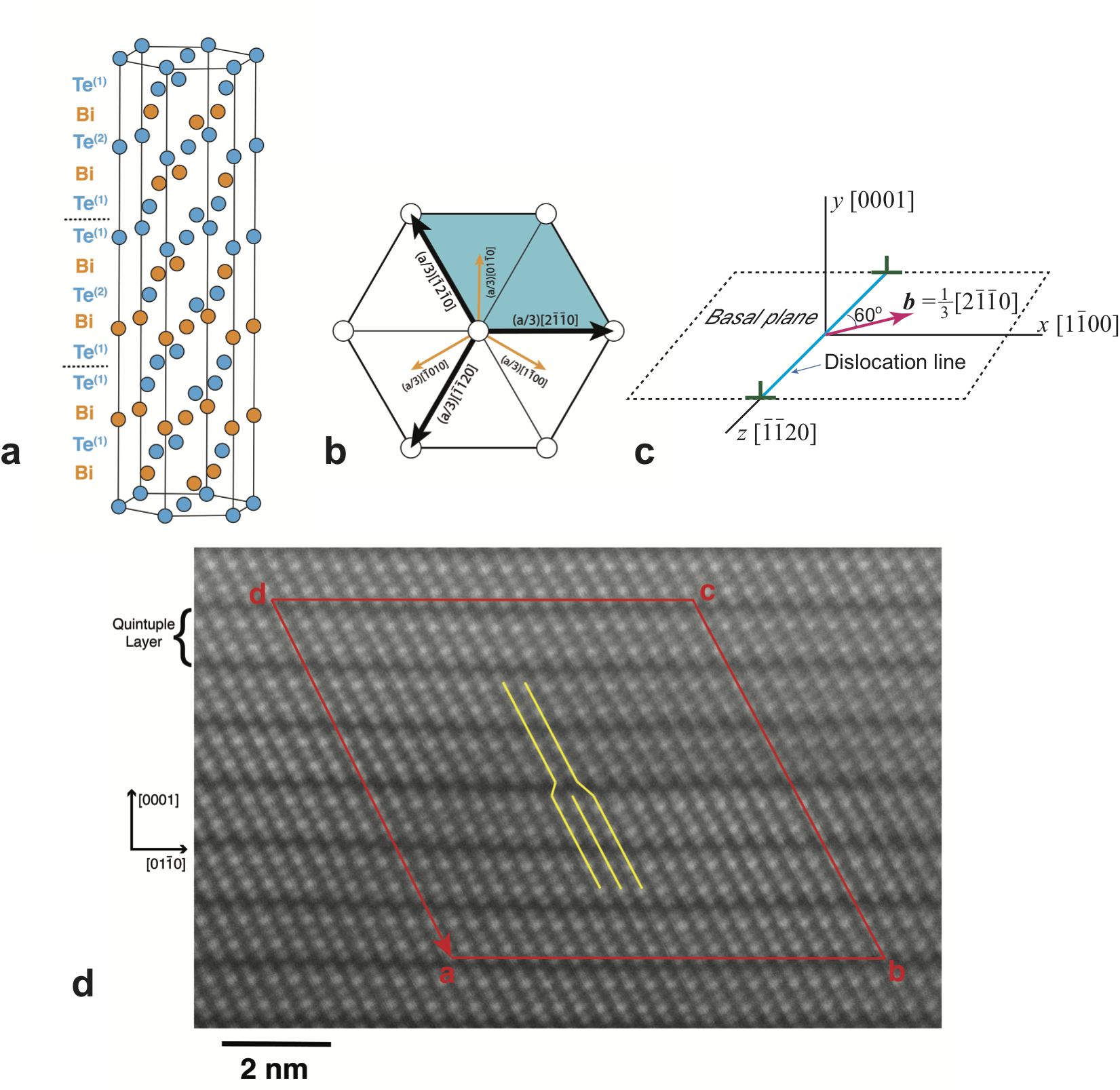}

\caption{Crystallographic details of \textcolor{black}{Bi$_{2}$Te$_{3}$}
and the dislocation. \textbf{a} Atomic arrangements
in Bi$_{2}$Te$_{3}$. The space between the Te$^{(1)}$:Te$^{(1)}$
planes is the van der Waals gap. \textbf{b} Projection of the structure
on the basal plane showing the Burgers vectors of dislocations. \textcolor{black}{The
unit cell is shaded in blue. The heavy black arrows show the Burgers
vectors for the $\frac{1}{3}\left\langle 2\bar{1}\bar{1}0\right\rangle $
type perfect lattice dislocations, whereas the smaller orange arrows
show the $\frac{1}{3}\left\langle 10\bar{1}0\right\rangle $ type
Burgers vectors that would result if Shockley partial dislocations
were to form.}  \textbf{ {c}} {{} Orientation of the
Cartesian axes relative to the dislocation line.}   \textbf{d} HRTEM image of Bi$_{2}$Te$_{3}$ projected
along $\left\langle 2\bar{1}\bar{1}0\right\rangle $ direction, showing
the quintuple layers and the Burgers circuit construction for calculation
of the dislocation Burgers vector \textcolor{black}{(see Supplementary
Note 2 for detail)}. The basal planes are horizontal and the yellow
lines trace $\left\{ 10\overline{1}5\right\} $ crystal planes one
of which terminates at the dislocation core. The scale bar represents
2 nm. \label{fig:Bi2Te3-structure}}
\end{figure}

\textbf{Experimental determination of the dislocation core structure.}
The crystal structure of Bi$_{2}$Te$_{3}$ is illustrated in Fig.~\ref{fig:Bi2Te3-structure}a.
The material has a rhombohedral crystal structure (R$\overline{3}$m,
$a=0.438$ nm, $c=3.05$ nm) \citep{RN266,RN421,RN267} consisting
of hexagonal sheets of alternating bismuth and tellurium atoms stacked
along the $c$-axis in 5-plane groupings, or quintuple layers, of
the form Te$^{(1)}$-Bi-Te$^{(2)}$-Bi-Te$^{(1)}$. Three of such
quintuple layers constitute a single unit cell. The van der Waals
gap is parallel to the basal plane and corresponds to the region between
abutting Te$^{(1)}$:Te$^{(1)}$ planes. The material is isomorphous
with Sb$_{2}$Te$_{3}$ and Bi$_{2}$Se$_{3}$, which are typically
alloyed with Bi$_{2}$Te$_{3}$ for thermoelectric applications \citep{RN158},
and which are of interest in their own right as topological insulators
\citep{RN597}. All these materials fall within the broader class
of tetradymite-type compounds \citep{RN13}, which possess similar
layered structures.

Previous work has identified two primary types of dislocation in Bi$_{2}$Te$_{3}$
and the isomorphous compounds: dislocations with Burgers vector, \textbf{b},
of type $\frac{1}{3}\left\langle 2\bar{1}\bar{1}0\right\rangle $
\citep{RN448,RN15,RN317,RN510,Fu:2017aa}, which lies parallel with
the basal plane \textcolor{black}{(Fig.~\ref{fig:Bi2Te3-structure}b,c)},
and those with \textbf{b} of type $\frac{1}{3}\left\langle 0\bar{1}11\right\rangle $
\citep{RN409,RN565}, which has a large component normal to the basal
plane. The Burgers vector \textbf{b} characterizes the magnitude and
direction of the lattice translation produced by the dislocation.
Operationally, it is given by the closure failure, when mapped onto
the perfect lattice, of an imaginary closed loop constructed around
the region including the dislocation (the so-called Burgers circuit)
\citep{Hirth}. We focus in this paper on the $\mathbf{b}=\frac{1}{3}\left\langle 2\bar{1}\bar{1}0\right\rangle $
type dislocations as their structure is directly related to the fundamental
interactions across the interlayer gap. Since this Burgers vector
is parallel to the basal plane, the dislocations can glide on this
plane under applied shear stresses.

The presence of $\frac{1}{3}\left\langle 2\bar{1}\bar{1}0\right\rangle $
dislocations in Bi$_{2}$Te$_{3}$ has long been recognized. In fact,
the pioneering electron microscopy study of $\frac{1}{3}\left\langle 2\bar{1}\bar{1}0\right\rangle $
dislocation networks in Bi$_{2}$Te$_{3}$ and Sb$_{2}$Te$_{3}$
by Amelinkx and Delavignette \citep{RN448,RN15}, conducted in the
early 1960s, provided one of the first direct observations of dislocations
in non-metallic materials. The assumption has long been that the cores
of such dislocations would be localized at the van der Waals gap between
the Te$^{(1)}$:Te$^{(1)}$ planes, where the bonding is weakest.
Indeed, \emph{in situ} observations of gliding dislocations in Bi$_{2}$Te$_{3}$-based
alloys provides some evidence supporting this assertion \citep{RN317,RN510}.
Recently, Fu et al.~\citep{Fu:2017aa} have shown that the cores
of $\frac{1}{3}\left\langle 2\bar{1}\bar{1}0\right\rangle $ dislocations
in Sb$_{2}$Te$_{3}$ are localized in the Te$^{(1)}$:Te$^{(1)}$
layers. However, direct proof of this localization in Bi$_{2}$Te$_{3}$
or detailed knowledge of the core structure of such dislocations have
been missing. \textcolor{black}{It was hypothesized \citep{RN15}
that the core was dissociated into $\frac{1}{3}\left\langle 01\bar{1}0\right\rangle $-type
Shockley partials by analogy with similar dissociation taking place
in close-packed metals. However, the microscope resolution at the
time did not permit the authors to test this hypothesis. As demonstrated
in the present paper, the suggested dissociation into partials does
not occur in Bi$_{2}$Te$_{3}$.}

To directly determine the core structure of the $\frac{1}{3}\left\langle 2\bar{1}\bar{1}0\right\rangle $
dislocations, we conducted an electron microscopic study of a polycrystalline
sample of Bi$_{2}$Te$_{3}$ consolidated through a thermomechanical
processes expected to produce dislocations. We employed the technique
of high angle annular dark-field (HAADF) scanning transmission electron
microscopy (STEM), using a probe-corrected instrument operated at
300 keV (see Methods for further details). This technique allows the
Te and Bi layers to be distinguished due to their large difference
in atomic number \citep{RN296}. Fig.~\ref{fig:Bi2Te3-structure}d
shows an atomically resolved HAADF-STEM image of a dislocation observed
in Bi$_{2}$Te$_{3}$. From Burgers circuit analysis, we confirm that
the dislocation is of $\frac{1}{3}\left\langle 2\bar{1}\bar{1}0\right\rangle $
type. The image itself is projected along a $\left\langle 2\bar{1}\bar{1}0\right\rangle $-type
orientation. Assuming that the line direction of the dislocation is
along this projection, the Burgers vector is inclined by $\pm60^{\circ}$
with respect to the dislocation line. Thus, the dislocation is of
$60^{\circ}$ mixed type with both edge and screw components. Inspection
of the image shows clearly that the dislocation core terminates at
the Te$^{(1)}$:Te$^{(1)}$ layer as expected. Note that the Bi$_{2}$Te$_{3}$
quintuples above and below the slip plane remain intact, suggesting
that the dislocation core maintains stoichiometry. We have analyzed
a total of 6 different dislocations of this type, all of which terminated
in the same manner. This can be contrasted to $\frac{1}{3}\left\langle 0\bar{1}11\right\rangle $
edge dislocations, which have been observed to form a dissociated,
Bi-rich core \citep{RN565}.

The question arises as to how localized the dislocation core is within
the glide plane. The degree of dissociation is absolutely central
to the properties and behavior of dislocations (e.g., controlling the
ease of deformation processes such as cross-slip). The dissociation
width depends on the shape of the so-called gamma-surface, which is
the excess interlayer energy $\gamma$ as a function of the translation
vector $\mathbf{t}$ parallel to the layers \citep{Vitek68,Vitek74,Balluffi95}.
The gamma-surface, in turn, is sensitive to the strength of the interlayer
interactions. Thus, measuring the core dissociation and reconstructing
the respective gamma-surface provides an effective way of assessing
the character of interlayer interactions.

To accomplish this, we measured the disregistry $\delta$ of the atomic
planes at the dislocation slip plane from the positions of $\left\{ 10\overline{1}5\right\} $
planes within the quintuple units above and below the slip plane.
The atomic plane positions on either side of the slip plane were determined
by the template averaging and matching method illustrated in Fig.~\ref{fig:Bi2Te3-TEM}a
and described in more detail in the Methods section.  {The
disregistry, $\delta=(u_{+}-u_{-})$, is calculated from the intersections
of the }$\left\{ 10\overline{1}5\right\} $ {{} planes
with the slip-plane. Here, $u_{+}$ is the intersection extrapolated
from the planes above the slip plane and $u_{-}$ is the intersection
extrapolated from below the slip plane.} The collection of disregistry
plots $\delta(x)$ measured from the 6 different dislocations is displayed
in Fig.~\ref{fig:Bi2Te3-TEM}b. The $x$-direction is parallel to
the edge component of the Burgers vector. The results were combined
into a single plot by binning along the $x$-axis and computing the
average and standard deviation of the disregistry in each bin (Fig.~\ref{fig:Bi2Te3-TEM}c).
\textcolor{black}{The striking feature of the disregistry plot is
that it shows a rather wide (over 1 nm) dislocation core yet no signs
of dissociation into well-defined partials, contrary to what was suggested
in \citep{RN15}.}

\smallskip{}
\smallskip{}

\noindent \textbf{Calculation of the dislocation core structure.}
To understand the nature of the unusually wide \textcolor{black}{and
yet undissociated} dislocation core, the gamma-surface on the basal
plane was computed by first-principles DFT methods. The challenge
of this calculation was that different exchange-correlation functionals
available in the literature produce qualitatively different shapes
of the gamma-surface, which in turn leads to different predictions
of the detailed core structure. To arrive at definitive conclusions,
several different DFT functionals were tested, including the local-density
approximation (LDA) \citep{kohn65}, the non-local correlation functionals
optPBE-vdW and optB88-vdW \citep{Dion:2004aa,Klimes:2010,Klimes:2011aa,Roman-Perez:2009aa}
accounting for dispersion interactions, and the DFT-D2 functional
\citep{Grimme:2006aa} introducing semi-empirical van der Waals corrections
(see Methods for details). The gamma-surface was computed with each
of these functionals (Fig.~\ref{fig:gamma}a). All functionals predict
the existence of a stable stacking fault (SF) on the basal plane,
which is obtained by a relative translation of the Te$^{(1)}$ layers
across the van der Waals gap. A stacking fault is a planar (2D) defect
of a crystal structure obtained by relative translation of two half-crystals
to create the wrong stacking sequence of the crystal planes. While
the translation vector corresponding to the local minimum of the fault
energy $\gamma_{\textrm{sf}}$ remains approximately the same, the
depth of the minimum and the respective SF energy depend on the functional.
The LDA approximation predicts a large SF energy and a shallow minimum,
whereas the DFT-D2 functional predicts the lowest SF energy with a
broad minimum surrounded by relatively high barriers.

Based on the computed gamma-surface $\gamma(\mathbf{t})$, the disregistry
function $\delta(x)$ in the dislocation core region can be predicted
theoretically using one of the Peierls-Nabarro type models \citep{Nabarro_book,Hirth}.
The classical Peierls-Nabarro model represents an edge dislocation
core by a continuous function $\delta(x)$ satisfying the boundary
condition $\delta(\infty)-\delta(-\infty)=b$, where $b$ is the magnitude
of the Burgers vector. It additionally postulates a sinusoidal shape
of the gamma surface in the $x$-direction, which is not suitable
for our purposes. Instead, we employed a semidiscrete variational
Peierls-Nabarro (SDVPN) model \citep{Bulatov1997,Lu00a,Lu2000b},
which was properly generalized in this work to capture the elastic
anisotropy of the Bi$_{2}$Te$_{3}$ crystal structure. In this model,
the disregistry $\delta_{i}$ is a vector (with three components labeled
by index $i$) and is evaluated at discrete values of $x$ equally
spaced by $\Delta x$. The latter corresponds to a spacing of atomic
columns parallel to the $x$-direction. The model can be applied to
an arbitrary (mixed-type) dislocation. The dislocation core is represented
by a discrete set of imaginary partial dislocations running parallel
to the actual dislocation line. The core energy is the sum of the
elastic strain energy arising from the interaction of the partial
dislocations, plus the crystal misfit energy due to the core spreading.
The latter contribution depends on the shape of the gamma-surface
provided as input to the model. In the variational formulation of
the model, the disregistry function $\delta_{i}(x)$ is found by minimizing
the total energy. Further details of the model are described in the
Supplementary Note 1.

\section*{Discussion}

\begin{figure}

\noindent \begin{centering}
\par\end{centering}
\includegraphics[width=0.99\textwidth]{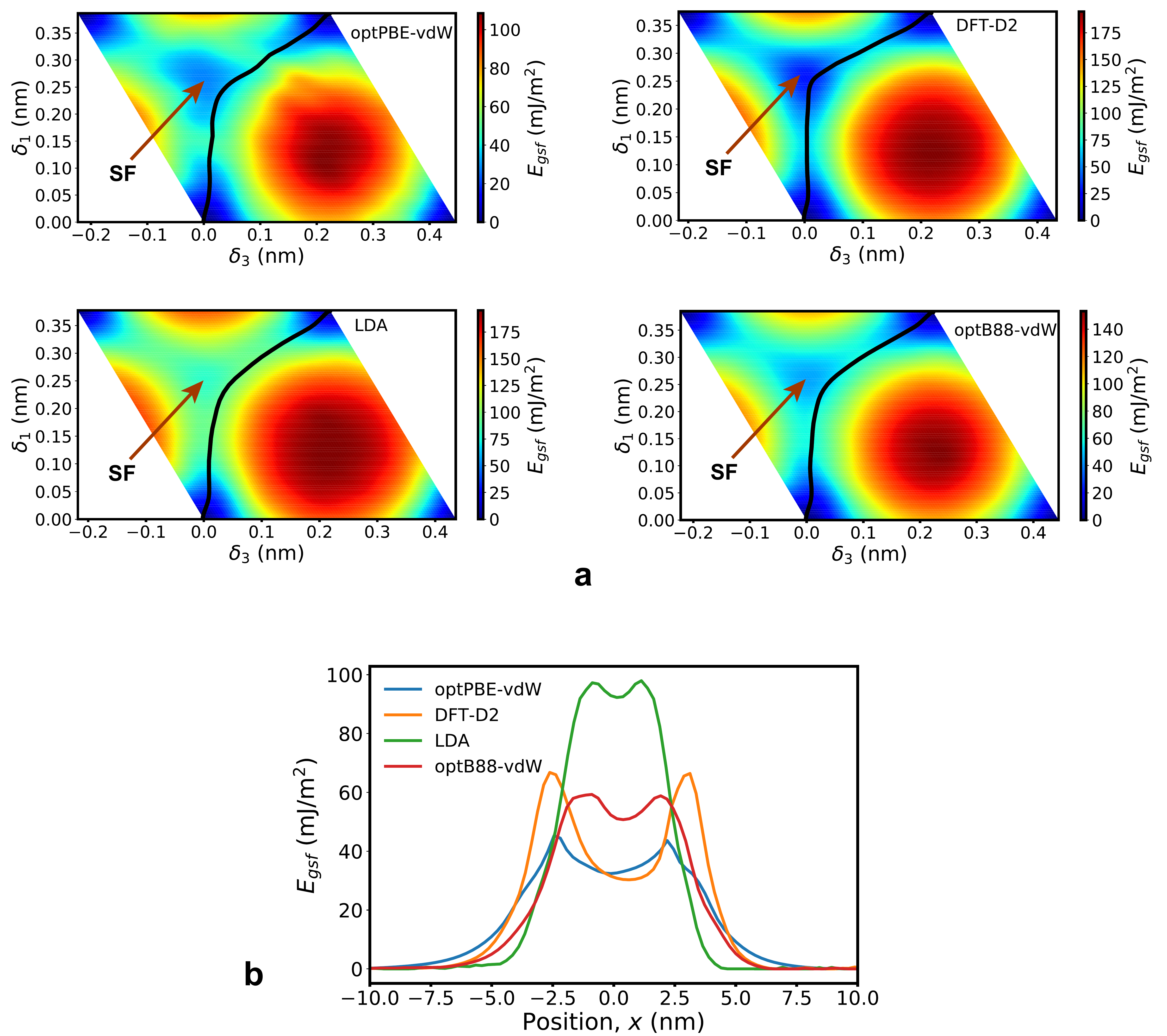}

\caption{The gamma-surface and the stacking fault. \textbf{a} Gamma-surfaces
computed with different DFT functionals. The generalized stacking
fault energy $E_{\textnormal{gsf}}$ is only shown within a repeat
unit parallel to the basal plane. The stable stacking fault position
(local minimum on the gamma-surface) is indicated. The black line
shows the disregistry path within the dislocation core region predicted
by the semidiscrete Peierls-Nabarro model using the respective gamma-surface.
\textbf{b} The generalized stacking fault energy as a function of
distance $x$ across the dislocation core computed with different
DFT functionals.\textcolor{black}{{} Note that, with the exception of
the DFT-D2 functional, all other DFT calculations predict remarkably
low barriers (on the order of 10 }mJ\,m$^{-2}$\textcolor{black}{)
on either side of the local minimum representing the stable SF.}}
\label{fig:gamma}

\end{figure}

Figure \ref{fig:Bi2Te3-TEM}c compares the SDVPN model predictions
with the experimental disregistry function. We emphasize that the SDVPN
calculations did not involve any fitting to experimental data. The
immediate conclusion is that the model predicts the core shape in
very reasonable agreement with experiment, confirming that the wide
spread of the dislocation core is primarily due to the weak van der
Waals bonding across the gap, not a result of dissociation into two
discrete partials separated by a SF ribbon as in low SF energy face-centered
cubic (FCC) metals. In the latter case, the disregistry function \textcolor{black}{would
have} a well-pronounced flat region at the center of the core, which
is not observed in Fig.~\ref{fig:Bi2Te3-TEM}c. This conclusion is
additionally supported by the disregistry trajectories (Fig.~\ref{fig:gamma}a),
showing the edge ($\delta_{1}$) and screw ($\delta_{3}$) components.
The trajectories pass close to the SF position but never reach it,
showing that a real SF ribbon does not form. (A similar situation
with very narrow dissociation without the formation of a real SF ribbon
is found in high SF energy FCC metals such as Al \citep{Woodward08a},
but then the dislocation core is much more compact than in Bi$_{2}$Te$_{3}$.)
A closer inspection of Fig.~\ref{fig:Bi2Te3-TEM}c reveals that the
different exchange-correlation functionals utilized for the DFT gamma-surface
calculations lead to slightly different shapes of the disregistry
plots. The LDA approximation underestimates the core width, which
is consistent with the prediction of the high SF energy (cf.~Fig.~\ref{fig:gamma}b).
The optPBE-vdW and especially DFT-D2 functionals create a kink at
the center of the core that is not supported by the experimental disregistry
function within the statistical error bars. This kink arises from
the relatively low SF energy and the wide separation of the energy
barriers around the energy minimum (cf.~Fig.~\ref{fig:gamma}b).
Thus, an additional outcome of the present calculations is that they
provide a useful benchmark of the DFT functionals, pointing to optB88-vdW
as an accurate model for van der Waals materials such as Bi$_{2}$Te$_{3}$.
The SF energy $\gamma_{\textrm{sf}}$ predicted by this functional
is 46 mJ\,m$^{-2}$ (Supplementary Table 1).

These results indicate that the wide spreading of the dislocation
core is primarily caused by the weak van der Waals bonding between
the quintuple layers\textcolor{black}{. This weak bonding results
in the relatively low unstable SF energy }$\gamma_{\textrm{us}}\approx60$\textcolor{black}{{}
}mJ\,m$^{-2}$\textcolor{black}{{} (the maxima in} Fig.~\ref{fig:gamma}b)\textcolor{black}{{}
compared to 65-200 }mJ\,m$^{-2}$\textcolor{black}{{} or higher in
FCC metals \citep{Li:2016aa,Bhogra:2014aa,VanSwygenhovenDF04,Bernstein:2004aa}.
In other words, the penalty for the translations of the layers parallel
to the van der Waals gap in Bi$_{2}$Te$_{3}$ is relatively small.
Perhaps more importantly, the depth $(\gamma_{\textrm{us}}-\gamma_{\textrm{sf}})$
of the local energy minimum corresponding to the stable SF in Bi$_{2}$Te$_{3}$
is unusually small, about 10 }mJ\,m$^{-2}$\textcolor{black}{. For
comparison, in Cu }$\gamma_{\textrm{sf}}$\textcolor{black}{{} is about
the same as in Bi$_{2}$Te$_{3}$ but }$\gamma_{\textrm{us}}$ is
about 160-180 mJ\,m$^{-2}$\textcolor{black}{{} \citep{Li:2016aa,Hunter:2013aa,Bernstein:2004aa}.
As a result, the stable SF minimum is surrounded by much higher }(about
100 mJ\,m$^{-2}$\textcolor{black}{) energy barriers. These high
barriers lead to the dissociation of the dislocation core structure
into cleanly separated, discrete }Shockley-partials. As another comparison,
the so-called MAX phases \citep{Barsoum:2000aa} also have a hexagonal
layered structure, but the bonding between the layers is not van der
Waals type. Unlike in Bi$_{2}$Te$_{3}$, the full dislocations on
the basal plane dissociate into discrete Shockley partials separated
by a stacking fault \citep{Higashi:2018aa}. The energy minimum corresponding
to the stacking fault is relatively shallow, but the absolute values
of all energies on the gamma surface are at least an order of magnitude
higher than in Bi$_{2}$Te$_{3}$, including the minimum depth \textcolor{black}{$(\gamma_{\textrm{us}}-\gamma_{\textrm{sf}})$}.
Similarly, in Ti the prismatic SF along the $\frac{1}{3}\left\langle 11\bar{2}0\right\rangle $
direction has a relatively shallow minimum, but the absolute values
of all energies are significantly higher than in Bi$_{2}$Te$_{3}$
\citep{Benoit:2013aa,Clouet:2015aa,Ready:2017aa}. The unique feature
of the basal dislocations in Bi$_{2}$Te$_{3}$, and most likely in
other tetradymite-structured chalcogenides, is the combination of
the low unstable SF energy with the shallowness of the stable SF minimum.
This combination \textcolor{black}{prevents any significant localization
of the dislocation content into partial dislocations. Even if such
dislocations hypothetically formed, their cores would be wide and
would strongly overlap, making the very concept of dissociation meaningless.
}Thus, despite the existence of a stable SF in the gamma surface,
the wide spreading of the dislocation core in Bi$_{2}$Te$_{3}$ is
\emph{not} \textcolor{black}{accompanied} by dissociation into two
discrete partials. The core remains wide but undissociated. This \textcolor{black}{important}
conclusion is supported by calculations within a semidiscrete Peierls-Nabarro
model with input from DFT calculations, showing excellent agreement
with experiment.

In summary, we have been able to resolve the detailed basal dislocation
core structure in the layered chalcogenide Bi$_{2}$Te$_{3}$, and
to quantify the atomic-plane disregistry across the dislocation core
using atomic-resolution electron microscopy. The wide spreading of
the dislocation core is mainly due to the weak, van der Waals type
bonding between the quintuple layers of the structure\textcolor{black}{,
which leads to relatively low energy barriers for the SF formation
and annihilation.} As an additional finding, we have identified the
exchange-correlation functional (optB88-vdW) that should be most appropriate
for future dislocation modeling in Bi$_{2}$Te$_{3}$ and, by extension,
other materials composed of van der Waals bonded atomic layers \citep{Tawfik:2018aa}.
As discussed in the beginning of the paper, the dislocation core structure
in Bi$_{2}$Te$_{3}$ can impact many functional properties of the
material. As additional evidence, Supplementary Note 3 reports on
preliminary DFT calculations of the electronic properties for an idealized
SF in Bi$_{2}$Te$_{3}$. The results suggest that the SF is likely
to affect the concentration of free charge carriers in the SF region.
A more detailed analysis of the dependence of the electronic structure
on the variation of the disregistry on the basal plane would be a
useful future step toward predicting the electronic effects of the
actual basal plane dislocations in Bi$_{2}$Te$_{3}$ and related
materials that exhibit a spread core.

\newpage\clearpage{}
\noindent \begin{center}
\textbf{METHODS}
\par\end{center}

\textbf{Experimental methodology.} The observations were conducted
on polycrystalline Bi$_{2}$Te$_{3}$ material that had been initially
consolidated by spark-plasma sintering and then further processed
by hot extrusion. This extrusion technique is convenient for the present
study since the deformation introduces dislocations and can drive
a preferential crystallographic texture of $\left\langle 2\bar{1}\bar{1}0\right\rangle $
\citep{Miura:2000aa}, which is an ideal imaging direction for atomically
resolving the dislocation core and its displacements.The sintering
employed procedures detailed in \citep{RN414}. Following sintering,
the Bi$_{2}$Te$_{3}$ puck was subjected to extrusion to increase
its density. During preparation, the puck was coated with a high temperature
graphite lubricant to ensure smooth movement through the tooling.
The tooling was heated to 400 $^{\circ}$C and held at this temperature
for 30 min to enhance Bi$_{2}$Te$_{3}$ plasticity during the extrusion
treatment. The puck was then pressed through a tool steel die with
a reduction ratio of 4:1 using a force of \textcolor{black}{$\sim10^{4}$} {{}
}\textcolor{black}{kg} at a strain rate of 0.1 s$^{-1}$.

The Bi$_{2}$Te$_{3}$ specimen for electron microscopy analysis was
thinned to electron transparency by mechanical grinding and dimpling,
followed by Ar$^{+}$ ion milling using a Fischione Model 1010 ion
milling system. The specimen was cooled during ion milling using a
liquid nitrogen stage. Analysis was conducted by HAADF STEM using
a probe-corrected 80-300 FEI Titan instrument operated at 300 keV.
Images were collected with the local grain region oriented along a
$\left\langle 2\bar{1}\bar{1}0\right\rangle $-type direction. In
order to efficiently identify dislocations over a large field of view,
initial imaging was conducted at low magnification with a scan sampling
frequency and scan orientation selected to provide strong Moiré contrast
from the dislocation. By imaging at low magnification, we also took
care to ensure that the analyzed dislocations were chosen to be well
separated from grain boundaries and other dislocations by a radius
of at least 75 nm. After identifying and focusing on each dislocation
from its Moiré image, higher magnification, atomic-resolution images
were collected. Image analysis was conducted using ImageJ and custom
routines written in MATLAB.

The disregistry across the dislocation slip plane was measured from
the atomically resolved HRSTEM images. A total of 6 dislocations were
analyzed from images with a sampling of 77 pixels/nm. Multiple images
were analyzed for each dislocation (4 images for one of the dislocations;
2 images each for the remaining 5 dislocations). We employed a template
averaging and matching approach \citep{RN584} to determine the atomic
plane positions on either side of the slip plane. We then computed
the difference in these positions as a function of distance to determine
the disregistry. An example illustrating the \textcolor{black}{template
matching approach} is shown in Fig.~\ref{fig:Bi2Te3-TEM}a. In detail,
the images were first digitally rotated to align the average $\left\{ 10\overline{1}5\right\} $
plane orientation parallel with the horizontal image axis. Next, an
initial, trial template region was defined by selecting a rectangular
region of the image encompassing a single Te$^{(1)}$-Bi-Te$^{(2)}$-Bi-Te$^{(1)}$
unit along a $\left\{ 10\overline{1}5\right\} $ plane. The trial
template was then cross-correlated with the starting image. Image
patches centered on the peaks in the cross-correlation function were
then averaged to give a refined template pattern. Finally, the refined
template pattern was cross-correlated with the starting image, and
the peaks from this function were extracted to determine the atomic
motif positions, and hence the local position $\left\{ 10\overline{1}5\right\} $
planes on either side of the slip plane.

The disregistry at the slip plane was determined by projecting the
$\left\{ 10\overline{1}5\right\} $ planes from above and below the
slip plane to the mid-plane between Te$^{(1)}$-Te$^{(1)}$ layers
and calculating the distance between these intersections. Since the
$\left\{ 10\overline{1}5\right\} $ planes are aligned parallel with
the $x$-axis of the analysis coordinate frame, the disregistry, $\delta=(u_{+}-u_{-})$,
was calculated for each plane intersection from the equation
\begin{equation}
u_{+}-u_{-}=(y_{+}-y_{-})\sqrt{\dfrac{1+m^{2}}{m^{2}}},\label{eq:1}
\end{equation}
where $y_{+}$ and $y_{-}$ are the $y$-coordinates of the motifs
measured on the two sides of the slip plane. $m$ is the slope of
the mid-plane line calculated by fitting lines to the motif positions
on the sides of the slip plane and taking the average slope for these
two lines.

\textbf{Simulation methodology.} First-principles DFT calculations
were performed using projector-augmented wave (PAW) pseudopotentials
as implemented in the electronic structure \textcolor{black}{Vienna
Ab initio Simulation Package (VASP)} \citep{Kresse1996,Kresse1999}.
Gamma-surface calculations were performed using 6-quintuplet slabs,
with changes in total energy obtained after translating the upper
3 quintuplets parallel to the hexagonal basal plane. The translation
vector $\mathbf{t}$ sampled the lateral area of the conventional
unit cell on a uniform $10\times10$ mesh. After each translation,
the atomic positions were relaxed in the direction normal to the fault.
The lateral lattice vectors \textcolor{black}{(in Cartesian coordinates)
$\mathbf{a}_{1}=a(1,0,0)$ and $\mathbf{a}_{2}=(a/2)(\sqrt{3},1,0)$}
and the initial atomic positions were obtained for the lattice parameter
$a$ computed by relaxing forces and stresses for bulk Bi$_{2}$Te$_{3}$.
Convergence with respect to the Brillouin zone sampling was achieved
employing uniform Monkhorst-Pack k-point meshes with sizes up to $7\times7\times1$.
A smooth function $\gamma(\mathbf{t})$ was obtained by interpolating
between the measured points using a multiquadric radial basis function.
Several exchange-correlation functionals were employed, namely, LDA
\citep{kohn65}, optPBE-vdW and optB88-vdW \citep{Dion:2004aa,Klimes:2010,Klimes:2011aa,Roman-Perez:2009aa},
and DFT-D2 \citep{Grimme:2006aa}. The lattice parameters and elastic
constants obtained with these functionals are summarized in the Supplementary
Table 1 and are in good agreement with previous DFT calculations \citep{Bjorkman:2012aa}.

In the SDVPN model used here, the disregistry function $\delta_{i}(x)$
was discretized on a mesh $\left\{ x^{\alpha}\right\} _{\alpha=1,...,N}$
of $N=300$ points with the spacing $\Delta x=a\sqrt{3}/3$. The total
dislocation energy is $E_{\textnormal{total}}=E_{\textrm{elastic}}+E_{\textrm{misfit}}$,
where the elastic strain energy $E_{\textrm{elastic}}$ depends on
the energy coefficients $K_{ij}$, which in turn depend on the elastic
constants $C_{ij}$. The misfit energy $E_{\textrm{misfit}}=\sum_{\alpha=1}^{N}\gamma\left[\delta_{i}(x^{\alpha})\right]\Delta x$
was obtained from the DFT gamma-surface. The equilibrium disregistry
$\delta_{i}(x^{\alpha})$ was found by numerical minimization of $E_{\textnormal{total}}$
under appropriate boundary conditions. As the initial guess we used
the arctangent disregistry function predicted by the classical Peierls-Nabarro
model. A continuous disregistry curve was obtained by smooth interpolation
between the $x^{\alpha}$ points. The calculations were performed
separately for each of the DFT functionals. Further technical details
of the SDVPN calculations can be found in the Supplementary Note 1.


\noindent \bigskip{}
 \bigskip{}

\noindent \textbf{Acknowledgements}

\noindent Y.~M.\ acknowledges support from the U.S. Department of
Energy, Office of Basic Energy Sciences, Division of Materials Sciences
and Engineering, the Physical Behavior of Materials Program, through
Grant No. DE-FG02-01ER45871.\textcolor{black}{{} The research at Sandia
National Laboratories was developed with funding from the Defense
Advanced Research Projects Agency (DARPA) (C.D.S.) and the Laboratory
Directed Research and Development program (D.L.M. and N.Y.). Sandia
National Laboratories is a multimission laboratory managed and operated
by National Technology \& Engineering Solutions of Sandia, LLC, a
wholly owned subsidiary of Honeywell International Inc., for the U.S.
Department of Energy's National Nuclear Security Administration under
contract DE-NA0003525. This paper describes objective technical results
and analysis. The views, opinions and/or findings expressed are those
of the authors and should not be interpreted as representing the official
views or policies of the Department of Defense, Department of Energy,
or the U.S. Government. Certain commercial instruments, materials
or processes are identified in this paper to adequately specify the
experimental procedure. Such identification does not imply recommendation
or endorsement by the National Institute of Standards and Technology,
nor does it imply that the instruments, materials or processes identified
are necessarily the best available for the purpose.}

\bigskip{}

\noindent \textbf{Author contributions}

\noindent D.\ L.\ M.\ performed the TEM observations of the Bi$_{2}$Te$_{3}$
dislocations and measurements of the disregistry, and prepared a draft
of the introduction and experimental part of the paper. N.\ Y.\ was
responsible for processing the Bi$_{2}$Te$_{3}$ material used in
this study. C.\ D.\ S.\ performed the DFT calculations and described
the results and methodology, while L.~M.~H.\ conducted the calculations
within the SDVPN model and prepared a draft of this part of the work.
D.\ L.\ M.\ and Y.\ M.\ conceived this project and coordinated
the collaborations among the co-authors. Y.~M.\ produced an initial
draft of the complete manuscript. All co-authors were engaged in discussions,
contributed ideas at all stages of the work, participated in the manuscript
editing, and approved its final version.

\bigskip{}

\noindent \textbf{Competing financial interests}

\noindent The authors declare no competing interests.

\bigskip{}

\noindent \textbf{Data availability}

\noindent Calculations within the SDVPN model were implemented in
Python and are openly available as part of the atomman Python package
at \url{https://github.com/usnistgov/atomman}. All data that support
the findings of this study are available in the Supplementary Information
file or from the corresponding author upon reasonable request.

\newpage\clearpage{}

\begin{figure}
\noindent \begin{centering}
\includegraphics[width=0.42\textwidth]{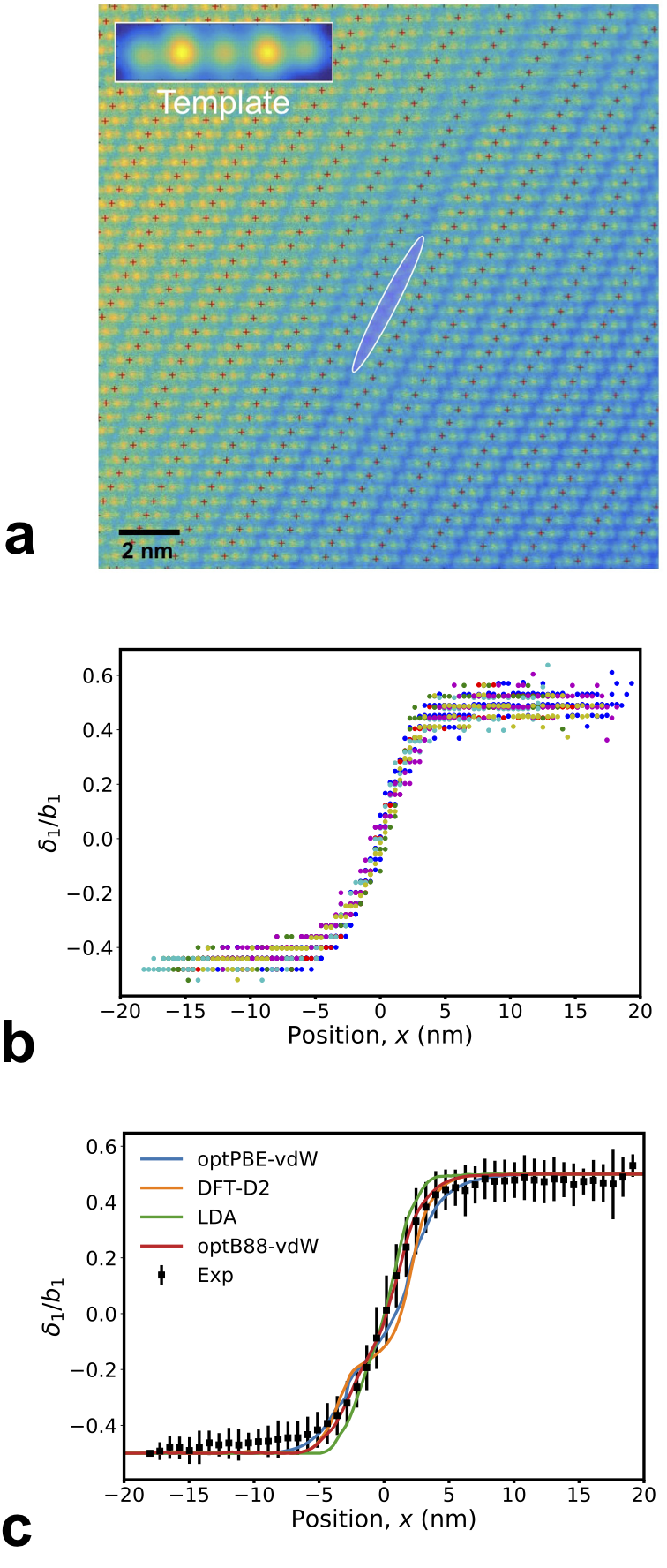}

\par\end{centering}
\caption{Disregistry at the dislocation core. \textbf{a} Illustration of the
templating procedure for extracting the atomic plane disregistry in
the dislocation core, with the cross symbols indicating the centers
of the quintuple structural units on either side of the slip plane.
The dislocation core is highlighted in purple. The scale bar represents
2 nm. \textbf{b} Disregistry as a function of distance $x$ across
the dislocation core for six (color-coded) dislocations observed in
this work. \textbf{c} Disregistry as a function of distance $x$ averaged
over the six dislocations and compared with predictions of the semidiscrete
Peierls-Nabarro model.\textcolor{black}{{} The experimental disregistry
$\delta_{1}$ has been normalized by the edge component $b_{1}$ of
the experimental Burgers vector.} The error bars represent two standard
deviations. The curves were obtained using different DFT functionals
indicated in the legend. \label{fig:Bi2Te3-TEM}}
\end{figure}

\end{document}